\documentclass[fleqn,twoside,twocolumn,nofootinbib]{revtex4} 
\usepackage[sec]{ujp}

\begin{document}
\title[Separate strangeness freeze-out]
{Separate chemical  freeze-out of strange particles  with conservation laws}

\author{D.R. Oliinychenko}
\affiliation{\bitp}
\address{\bitpaddr}
\email{dimafopf@gmail.com}
\affiliation{FIAS, Goethe-University, Frankfurt}
\address{Ruth-Moufang Str. 1, 60438 Frankfurt upon Main, Germany}

\author{V.V. Sagun}
\affiliation{\bitp}
\address{\bitpaddr}
\email{v_sagun@ukr.net}

\author{A.I. Ivanytskyi}
\affiliation{\bitp}
\address{\bitpaddr}
\email{a_iv_@ukr.net}

\author{K.A. Bugaev}
\affiliation{\bitp}
\address{\bitpaddr}
\email{bugaev@th.physik.uni-frankfurt.de}

 \udk{539.12} \pacs{25.75.-q, 25.75.Nq} \razd{\seci}

\setcounter{page}{1}%
\maketitle

\begin{abstract} 
The
Hadron Resonance Gas Model  with two chemical freeze-outs, connected by conservation laws is considered. 
We are arguing that the chemical freeze-out of strange hadrons should occur earlier than the chemical freeze-out  of non-strange hadrons.
The hadron multiplicities measured in the heavy ion collisions for  the center of mass  energy range 2.7 - 200 GeV are described well by such a model. Based  on a success of such  an approach, a radical way to improve the Hadron Resonance Gas Model  performance is suggested. Thus, we suggest to identify the  hadronic reactions that freeze-out noticeably earlier or later that most of the others reactions (for different collision energies they may be different) and to consider  a separate freeze-out for them. 
\end{abstract}



\section{Introduction}

The hadronic  multiplicities  measured in heavy ion collisions and in the collisions of elementary particles are traditionally described by the Hadron Resonance Gas Model (HRGM) \cite{Thermal_model_review, KABAndronic:05,ref:Beccattini97,Becattini:gammaHIC, KABugaev:Horn2013}. 
Its is based on an assumption  that the fireballs produced in such collisions reach a full thermal equilibrium. 
Using this assumption it is possible 
to describe the hadronic multiplicities  registered in experiment with the help of two parameters:  temperature $T$ and baryo-chemical potential $\mu_B$. Parameters $T$ and $\mu_B$ obtained from the  fit of multiplicities for different collision energies correspond to  the stage  of chemical freeze-out. 
Its physical meaning is that at this stage  the inelastic collisions cease simultaneously for all sorts of particles.
However, in such a simple form the concept of chemical freeze-out  works well for the hadrons which consists of the $u$ and $d$ (anti)quarks, while the  strange hadrons demonstrate deviation from chemical equilibrium. At the same  time 
the hydrodynamic simulations (see e.g. a  review \cite{Hydro_rev}) rather successfully reproduce 
the transverse momentum spectra of strange particles. This is an old  problem of  the thermal approach   and 
in order to account for  an  observed  deviation  of  strange particles from the complete chemical equilibrium the additional parameter $\gamma_s$, the strangeness suppression factor, was suggested  
\cite{Rafelski:gamma} long ago. 
Although the concept of strangeness suppression proved to be important both in the collisions of elementary particles \cite{Becattini:gammaHIC} and in nucleus-nucleus collisions \cite{Becattini:gammaHIC, PBM:gamma} the problem of its justification remains unsolved. 
Thus, up to now it is unclear what is the main physical reason which is responsible for  chemical non-equilibrium  of strange hadrons.   

Moreover, it is well known  \cite{KABAndronic:05} that  the fit of hadron multiplicities with the strangeness suppression factor $\gamma_s$ improves the quality  of data description, but  still the  fit seldom attains a good quality, especially  at low collision energies. 
This is clearly seen from the center of mass energy  behavior of two most prominent  ratios 
that involve the lightest strange meson, i.e. $K^+/\pi^+$, and the  lightest strange baryon, i.e. $\Lambda/\pi^-$,
which, so far, cannot be successfully reproduced  \cite{Becattini:gammaHIC, KABAndronic:05, PBM:gamma} by the traditional versions of the HRGM. 
Also  the ratios involving the  multi-strange hyperons  $\Xi$ and $\Omega$  exhibit 
an apparent failure  of the $\gamma_s$ fit at the center of mass energy $\sqrt{S_{NN}} = 8.76,  12.3$ and 17.3 GeV \cite{KABAndronic:05}. 
Since  the $\gamma_s$ fit does not 
improve their description sizably, we conclude that there should exist a different reason for the apparent deviation of strange hadrons from chemical equilibrium and, hence, the concept of chemical freeze-out requires a further development.

Recently  an alternative concept of chemical freeze-out  of strange hadrons was suggested \cite{BugaevEPL:13}.
Instead of a simultaneous  chemical freeze-out  for  all hadrons the  two different chemical freeze-outs were suggested: one for  particles, containing strange charge, even hidden, (we refer to it as strangeness freeze-out, i.e. SFO) and another one (FO) for all other hadrons which contains only $u$ and $d$ (anti)quarks. 
A partial  justification for the  SFO hypothesis 
 is given in  \cite{EarlyFO:1,EarlyFO:2,EarlyFO:3}, where the early chemical and kinetic FO of  $\Omega$ hyperons and $J/\psi$ and $\phi$ mesons is discussed for the energies at and  above the highest SPS energy. 
In this  article we further develop and refine  the  SFO concept  of Ref.  \cite{BugaevEPL:13}, and present here 
a more coherent and  detailed   picture of two freeze-outs together with   new arguments which allow us to better  justify and to  improve the  performance of the  HRGM.

The paper is  organized as follows. In the next section we discuss the concept of chemical freeze-out in some details and give the arguments that in a meson dominated hadronic medium the SFO should occur earlier than the FO. Section 3 is devoted to a description of the HRGM with the multicomponent hard-core repulsion. The results are presented in Section 4, while Section 5 contains our conclusions and suggestions. 

\section{The  Framework of  Thermal Model}

In 1950 in his pioneering paper \cite{ref:Fermi50} E. Fermi suggested to use the statistical model to find the outcome of high energy nucleon-nucleon collisions. Since in such reactions there were produced from 10 to 30 hadrons, they were named as the processes of  multihadron production. According to E. Fermi, the large number of particles in a finale state of these processes  naturally suggested to apply the methods of statistical mechanics. 
The next crucial step suggested by  E. Fermi  was a justification of  the  thermal equilibrium assumption due to strong interaction between the  particles.
A few years later L. D. Landau suggested to apply the relativistic hydrodynamics to the reactions of  multihadron production \cite{ref:Landau53}, because   the applicability  conditions of  relativistic hydrodynamics
are  basically the same as  for 
 the full  (local) thermal equilibrium, if  the strong discontinuities are absent. 

Since that time an  assumption of thermal equilibrium at some stage of the multihadron production reactions   was tested experimentally both in the nucleon-nucleon collisions   and in the collisions of heavy ions. In other words, the  outcome of  such reactions was compared to the results of statistical models. 
A coincidence between the statistical models predictions and the experimental results appeared to be good both for the nucleon-nucleon collisions and for the heavy ion collisions at energy range starting from the center of mass energy  $\sqrt{S_{NN}} = 2$ GeV per nucleon in the fixed target experiments performed  at the Brookheaven AGS up to the center of mass energy  $\sqrt{S_{NN}} = 2.76$ TeV  achieved at the Large Hadron Collider \cite{KABAndronic:05, ref:Andronic_LHC}.  It was even suggested that for the high energy electron-positron collisions the statistical model can also describe  the hadron multiplicities \cite{ref:Beccattini97}.  However,  later 
on a more thorough analysis \cite{ref:Andronic_ee} showed that even within rather sophisticated canonical ensemble consideration the  discrepancy  between theory and experiment is rather large with $\chi^2/dof > 5$. 

Now let us consider in some details  a particular set of models used to describe hadron multiplicities in nucleon or heavy ion collisions, that are known as the HRGM \cite{Thermal_model_review, KABAndronic:05,ref:Beccattini97,Becattini:gammaHIC, KABugaev:Horn2013, PBM:gamma,ref:Andronic_LHC}. A common feature of this set of models is an assumption that at some moment there exists  a fireball consisting of all possible hadronic states being  locally  in  thermal and chemical equilibrium. The term chemical equilibrium  means that rates of forward and backward reactions are equal, i.e. for any hadron specie the rate of its production is equal to the rate of its destruction. The characteristic time  
of equilibration varies with collision energy, but one can safely say that it lies within  the interval of 0.1-10  fm/c \cite{Equilirb94, Equilirb95,Equilirb95b}. This means that one can safely ignore  weak interaction, because its characteristic time is essentially longer. Therefore, the baryon charge $B$, the strange charge $S$, the isospin projection $I_3$, the charm charge  $C$ and  the bottom charge  are conserved in almost all hadron reactions. Some of the most frequent hadronic reactions reactions read: $\pi\pi \to \rho \to \pi \pi$, $\pi K \to K^* \to \pi K$, $\pi N \to \Delta \to \pi N$. They lead to thermal equilibration, but do not change  the number of particles. Another reactions, such as $\pi N \to N^* \to \Delta \pi \to N \pi \pi$, change the number of particles and lead to the chemical equilibration. Was such a system of all hadron states kept in a finite box of volume $V$, it would inevitably equilibrate both thermally and chemically at $t \to \infty$. Let us define the  characteristic time of equilibration between the species A and B  $\tau_{AB}$ as an average time when $N_0$ collisions between A and B occurred. If there are only  A and B species in the box then $\tau_{AB} \sim \frac{1}{n_A n_B \sigma_{AB}}$, where $\sigma_{AB}$ is a cross-section of AB reaction and $n_{A} (n_{B})$ denote the  concentration of   specie A (B). If one considers a gas of many species in the box out of equilibrium, then the equilibration times will be defined from the system of equations (assuming  only the  reactions $2 \to 1$ and $1 \to 2$):
\begin{eqnarray}
\frac{dN_i}{dt} &=& \sum_{AB} \frac{N_A N_B v_{AB}^{rel}}{V}  \sigma_{AB \to i} -
                    \sum_{A}  \frac{N_i N_A v_{Ai}^{rel}}{V}  \sigma_{Ai \to B} - \nonumber \\
               & & -\sum_{CD} \Gamma_{i\to CD} N_i\,,
\label{eq:equil}
\end{eqnarray}
where $N_i$, $N_A$, $N_B$ are  the number of hadrons of corresponding   kind, $\sigma$ denotes the corresponding cross-sections and $\Gamma$ is the decay rate. The first term on the right hand  side describes the formation of particles of kind $i$, the second term  stands for the  particle destruction of this kind  in the  $2 \to 1$ reaction and the third term stands for the decays of this kind of particles. From these  equations one can see that the larger production cross-section leads to a faster equilibration, while the  larger volume leads to a slower equilibration. One can also see that depending on cross-sections of production and decay and also on volume, equilibration times for different species may be different. These equations are, of course,  oversimplified, because they do not include the momentum dependencies. If one introduces such dependencies, then  one obtains  the system of  Boltzmann equations, and, hence,  Eq. (\ref{eq:equil}) can be regarded as the system of  Boltzmann equations averaged over  momenta. 
However, even  these oversimplified equations can help to understand the way how a system approaches an equilibrium. For instance, from Eq. (\ref{eq:equil})   one can see that increasing the box volume $n$ times is equivalent to decreasing all the cross-sections $n$ times. One can also see that for very large volumes  only the decays will occur. 

If the system is expanding, i.e.  $V=V(t)$,  then there is no guarantee that all particle species will be at chemical and thermal equilibrium at any time. The  simplest way to qualitatively characterize an expanding system  is to introduce a set of characteristic times: expansion time $t_{ex}$, thermalization time $t_{th}$ and chemical equilibration $t_{ch}$ time for different species. It is known that typically for the reactions of strongly interacting particles there is an inequality   $t_{ch} \gg t_{th}$ \cite{Equilirb94,Equilirb95,Equilirb95b}. It is equivalent to a statement that cross-sections of reactions which  lead to a chemical equilibration are much  smaller than  the cross-sections of reactions which  lead to a thermalization. During the expansion process the system  volume increases or equivalently  one can say that all cross-sections effectively decrease  in the same factor. Therefore, the reactions which  lead to a chemical equilibration will cease earlier, than 
the reactions which  lead to a thermalization and they, respectively,  are called as  chemical and kinetic 
 freeze-out. Since the cross-sections of different reactions are not the same, generally one can talk about chemical and kinetic freeze-out for  each particle specie. 

Typically in vacuum the reactions involving  strange particles  have smaller cross-sections than the reactions involving only non-strange particles (charm and bottom are not considered here at all). Then from  our previous consideration one can conclude that, if  the cross-sections and the  thresholds of hadronic reactions occurring at the  late stage of expansion  do not differ from their vacuum values, then   the chemical equilibrium for strange particles should be  lost earlier.  The kinetic freeze-out for strange particles is also going to occur  earlier than the kinetic freeze-out of non-strange hadrons,  but  later than the chemical freeze-out for any hadron specie. 
These conclusions are based on  the following hierarchy of the switching off times of   hadronic reactions:
\begin{eqnarray}
t_{K \Lambda \to \Sigma p } >  t_{\pi N \to N^* \to \Delta \pi \to N \pi \pi} \gg \nonumber\\
\gg t_{K \pi \to  K^* \to K \pi} > t_{N \pi \to \Delta \to N \pi} \,. 
\end{eqnarray}\label{EqII}
%
It is not only cross-sections that influence  the freeze-out times. As one can see from Eq. (\ref{eq:equil}), the smaller concentrations are, the lower rate of reactions is expected.  The numbers of strange particles different from kaons are smaller than the number of protons, and  this is one more factor that makes 
slower the reactions of strangeness exchange  and leads to an  earlier freeze-outs  of strange particles.
Of course, one should keep  in mind that this simplified treatment is valid at low particle densities, if  an approximation of binary reactions is reasonable and if the surrounding medium does not essentially  modify the reaction threshold. Therefore, appearing of the results that contradict to the conclusions above 
should be considered as a signal that the chemical freeze-out picture based on Eqs. (\ref{eq:equil}) and (\ref{EqII}) is not justified and, hence,  one has to seek for another explanation. 

Nevertheless, the argumentation above motivates  to consider  a separate chemical freeze-out of strange particles  in the HRGM. This was done recently in two independent studies   \cite{BugaevEPL:13, BugaevNICA:13} and \cite{ref:Gupta13}. In \cite{ref:Gupta13} three free parameters were taken for FO  (temperature, baryon chemical potential and volume) and three free parameters of the same kind for SFO. The electric charge chemical potential $\mu_Q$ was taken from the condition $N_B/N_Q$ = 2.5 for both freeze-outs. Species subjected to the SFO  were all strange particles and the  $\phi$ - mesons. 
The strange charge  was treated canonically and the particle multiplicities were fitted. An approach of \cite{BugaevEPL:13,BugaevNICA:13} is quite different. The parameters of FO and SFO  were connected by the conservation laws, namely the baryon number conservation, the $I_3$ conservation and the  entropy conservation. Both freeze-outs were treated grand canonically and the $\phi$  mesons were not subjected to earlier freeze-out. Also, in contrast to oversimplified treatment of the equation of state,  the HRGM of \cite{BugaevEPL:13} includes the width of all hadron resonances and the  short range repulsion which is taken into account via the excluded volume corrections, while in \cite{ref:Gupta13} these important features  are neglected. 

We would like to stress, although being simple and successful in describing the hadronic multiplicities, an approach suggested in  \cite{ref:Gupta13} violates the above mentioned conservation laws. Moreover, in such approach it might happen that not only the entropy conservation is violated, but  entropy may decrease  from an earlier freeze-out to the later one. Finally, while the number of fitted multiplicities is rarely exceeding 10 per  one collision energy value, having six fitting   parameters for each energy value seems to be excessive. Therefore, below we outline an alternative model \cite{BugaevEPL:13}, which seems to be physically more relevant.

\section{Model formulation}

In the simplest version  the HGRM represents   the   gas of hadrons being in chemical and thermal equilibrium which is described by the grand canonical partition function. The  multiplicity of particles of 
the mass $m_i$ and degeneracy $g_i$ is  given by:
\begin{eqnarray}
N_{i} = g_i V \int \frac{d^3k}{(2\pi)^3} \frac{1}{e^{\sqrt{m_i^2 + k^2}/T - \mu} \pm 1}\,,
\end{eqnarray}
where the sign $+ (-)$ in the  equation above  stays for Fermi (Bose) statistics and $\mu_i$ denotes 
 the full  chemical potential $\mu_i = \mu_B B_i + \mu_S S_i + \mu_{I3} I_{3i}$ of particles of sort $i$,
 $B_i $ is their baryonic charge, $S_i$ is their strange charge and $I_{3i}$ denotes their  third projection of isospin.
The 
chemical potentials $\mu_B$, $\mu_S$ and $\mu_{I3}$ which correspond to the conserved charges  can be found from the conservation laws
\begin{eqnarray}
\sum\limits_{i} N_i B_i &=& B^{init}\,, \\
\sum\limits_{i} N_i S_i &=& S^{init} \,, \\
\sum\limits_{i} N_i I_{3_i} &=& I_{3}^{init}\,,
\end{eqnarray}
then the temperature $T$ and the system volume $V$ will be free parameters. One can, however, take $T$ and $\mu_B$ as free parameters and this is a conventional choice. In \cite{ref:Oliinychenko13} we argued that for midrapidity the quantities  $B^{init}$ and $I_{3}^{init}$ are  anyway unknown, so one can fit the ratios and have $T$, $\mu_B$ and $\mu_{I3}$ as the fitting  parameters. Using this procedure one gets 
the hadron multiplicities that correspond to the full  thermal
equilibrium. To get  the final particle multiplicities, one has to take  into account the decays of hadron resonances (see below).

An extension of the HRGM to two freeze-outs is almost obvious in the case of \cite{ref:Gupta13}, where both non-strange and strange freeze-outs have their own parameters and are by no means connected. In such a case one considers two separate ideal gases with their own parameters. 
However, if one follows the way described in \cite{BugaevEPL:13}, then some complications arise. One problem is to properly include the conservation laws, then one has  to take the excluded volume into account in a consistent way. By consistency we mean that the standard  thermodynamic identities should be obeyed. One more issue is  the change of entropy between two freeze-outs due to decays of strange resonances. However, as  we  argued in \cite{BugaevEPL:13}  the latter is negligible, because the time interval between  two freeze-outs is short.

Also we would like to stress  that the excluded volume for all particles remains the same after the SFO. Indeed, not all reactions between the strange and non-strange particles cease, but only those with the strangeness exchange. For instance, the reaction $\pi K \to K^* \to \pi K$ survives  after the SFO. It keeps the same excluded volume between pions and kaons, but does not provide the chemical equilibrium for kaons. 

After these comments let us formulate our  approach. 
It is based on the multicomponent formulation of the HRGM \cite{KABugaev:Horn2013}, which is currently the best at describing the observed hadronic multiplicities. 
Therefore, it is natural  to apply such a formulation to describe 
 both the FO  and the  SFO.
 The present  HRGM  was worked out  in \cite{KABugaev:Horn2013,ref:Oliinychenko13,RVDW:1,MultiComp:08,RVDW:2,MultiComp:13b,Bugaev:1312a,Bugaev_SFO_13}.
The interaction between  hadrons  is taken into account via the hard-core radii, with the different  values for pions $R_{\pi}$, kaons $R_K$, other mesons $R_m$ and baryons $R_b$. The best fit values for such radii $R_b$ = 0.2 fm, $R_m$ = 0.4 fm, $R_{\pi}$ = 0.1 fm, $R_K$ = 0.38 fm were obtained in \cite{KABugaev:Horn2013}. The main equations of the model are listed below, but more details of the model can be found in \cite{KABugaev:Horn2013,ref:Oliinychenko13}.

We consider the Boltzmann gas of $N$ hadron species in a volume $V$ that has  the temperature $T$, the baryonic chemical potential $\mu_B$, the  strange chemical potential $\mu_S$ and the chemical potential of the isospin third component $\mu_{I3}$. The system  pressure $p$ and the $K$-th charge density $n^K_i$ ($K\in\{B,S, I3\}$) of the i-th hadron sort are given by the expressions  
\begin{eqnarray}\label{EqIn}
%
%
\label{EqIIn}
\frac{p}{T} =  \sum_{i=1}^N \xi_i \,, ~~n^K_i =\frac{ Q_i^K{\xi_i}}{\textstyle  1+\frac{\xi^T {\cal B}\xi}{\sum\limits_{j=1}^N \xi_j}}, ~~\xi  = \left(
\begin{array}{c}
 \xi_1 \\
 \xi_2 \\
 ... \\
 \xi_N
\end{array}
\right), 
%
\end{eqnarray}
where $\cal B$ denotes a symmetric  matrix of the second  virial coefficients with the elements $b_{ij} = \frac{2\pi}{3}(R_i+R_j)^3$ and 
 the variables $\xi_i$ are the solutions of the following system
\begin{eqnarray}\label{EqIII}
&&\hspace*{-4mm}\xi_i =\phi_i (T)\,   \exp\Biggl[ \frac{\mu_i}{T} - {\textstyle \sum\limits_{j=1}^N} 2\xi_j b_{ij}+{\xi^T{\cal B}\xi} {\textstyle \left[ \sum\limits_{j=1}^N\xi_j\right]^{-1}} \Biggr] \,, \quad \quad \\
&&\hspace*{-4mm}\phi_i (T)  = \frac{g_i}{(2\pi)^3}\int \exp\left(-\frac{\sqrt{k^2+m_i^2}}{T} \right)d^3k  \,.
\label{EqIV}
\end{eqnarray}
Here   the full chemical potential of the $i$-th hadron sort is defined  as before,  $ \phi_i (T) $ denotes 
the thermal particle  density of  the $i$-th hadron sort of mass $m_i$ and degeneracy $g_i$, and  $\xi^T$  denotes  the row of  variables $\xi_i$.  

The width correction is taken into account by averaging all expressions containing resonance mass by the Breit-Wigner distribution having a  threshold (see, for instance \cite{Thermal_model_review}, for more details). The effect of resonance decay $Y \to X$ with the branching ratio $BR(Y \to X)$  on the final hadronic multiplicity is taken into account as $n^{fin}(X) = \sum_Y BR(Y \to X) n^{th}(Y)$, where $BR(X \to X)$ = 1 for the sake of convenience. The masses, the  widths and the strong decay branchings of all hadrons  were  taken from the particle tables  used  by  the  thermodynamic code THERMUS \cite{THERMUS}.

The SFO  is assumed to occur for all strange particles  at the temperature $T_{SFO}$, the baryonic chemical potential $\mu_{B_{SFO}}$, the isospin third projection chemical potential $\mu_{I3_{SFO}}$ and the three dimensional space-time extent (effective volume) of the freeze-out hypersurface  $V_{SFO}$. The FO of   hadrons which are built of the $u$ and $d$ (anti)quarks,  is assumed to be described by  its own  parameters $T_{FO}$, $\mu_{B_{FO}}$, $\mu_{I3_{FO}}$, $V_{FO}$. Eqs. (\ref{EqIn})--(\ref{EqIV}) for FO and SFO remain the same as for a simultaneous  FO of all particles.
In both cases $\mu_S$ is found from the net zero strangeness condition. 
The major difference of the SFO approach 
 is the presence of   conservation laws and the corresponding modification  of  multiplicities due to resonance decays. 
Thus, we assume that between two freeze-outs  the system is sufficiently dilute
and hence its evolution is governed by the continuous hydrodynamic evolution which conserves 
the entropy. Then 
equations for  the entropy, the baryon charge and  the isospin projection conservation connecting two freeze-outs are as follows:
\begin{eqnarray}
s_{FO} V_{FO} = s_{SFO} V_{SFO} \,, \label{ent_cons}\\
n^B_{FO} V_{FO} = n^B_{SFO} V_{SFO} \,, \label{B_cons}\\
n^{I_3}_{FO} V_{FO} = n^{I_3}_{SFO} V_{SFO} \,. \label{I3_cons}
\end{eqnarray}
Getting rid of the effective  volumes we obtain
\begin{eqnarray} 
\label{Eq:FO_SFO1}
\frac{s}{n^B} \biggl|_{FO} = \frac{s}{n^B} \biggr|_{SFO} \,,  \quad  \frac{n^B}{n^{I_3}} \biggl|_{FO} = \frac{n^B}{n^{I_3}} \biggr|_{SFO}
%
 \,.  
\end{eqnarray}
Therefore, the variables $\mu_{B_{SFO}}$ and $\mu_{I3_{SFO}}$ are not free parameters, since
 they are found from the system  (\ref{Eq:FO_SFO1})  and only $T_{SFO}$ should be fitted.
Thus, for the SFO the number of independent fitting parameters is 4 for each value of collision energy. 

The number of resonances  appeared due to decays are found  from 
\begin{eqnarray} \label{Eq:SFO_decays}
%
\frac{N^{fin}(X)}{V_{FO}} = \sum_{Y \in FO} BR(Y \to X) n^{th}(Y) + \nonumber \\
 \sum_{Y \in SFO} BR(Y \to X) n^{th}(Y) \frac{V_{SFO}}{V_{FO}} \,. 
\end{eqnarray}
Technically this is done by multiplying all the thermal concentrations for SFO by $n^B_{FO}/n^B_{SFO} = V_{SFO}/V_{FO}$ and applying the conventional resonance decays. 

\section{Results}

{\bf Data sets and fit procedure.} In our choice of the data sets we basically  followed Ref. \cite{KABAndronic:05}. Thus, at  the AGS energy range of collisions ($\sqrt{S_{NN}} = 2.7 -4.9$ GeV) the data are  available for the kinetic beam energies from 2 to 10.7 AGeV.  For the beam energies 2, 4, 6 and 8 AGeV there are only a few data points available: the yields for pions \cite{AGS_pi1, AGS_pi2}, for protons \cite{AGS_p1, AGS_p2}, for kaons  \cite{AGS_pi2} (except for 2 AGeV), for  $\Lambda$ hyperons the integrated over $4 \pi$ data are available \cite{AGS_L}. For the beam  energy 6 AGeV there exist  the $\Xi^-$ hyperon  data integrated over $4 \pi$ geometry   \cite{AGS_Kas}. However, the data for  the  $\Lambda$ and $\Xi^-$ hyperons have to be corrected \cite{KABAndronic:05}, and instead of the raw experimental data we used their  corrected values of  Ref. \cite{KABAndronic:05}. For the highest AGS center of mass energy $\sqrt{S_{NN}} = 4.9$ GeV (or the beam energy 10.7 AGeV) in addition to the mentioned data for pions, (anti)protons and  kaons  there exist data for $\phi$ meson \cite{AGS_phi}, for  $\Lambda$ hyperon \cite{AGS_L2} and for   $\bar \Lambda$ hyperon \cite{AGS_L3}. Similarly to \cite{KABugaev:Horn2013}, here  we analyzed  only  the  NA49  mid-rapidity data  \cite{KABNA49:17a,KABNA49:17b,KABNA49:17Ha,KABNA49:17Hb,KABNA49:17Hc,KABNA49:17phi} since they are traditionally the most difficult to describe. Because   the RHIC high energy  data of different collaborations agree with each other, we  present the analysis of  the STAR results  for $\sqrt{S_{NN}}= 9.2$ GeV \cite{KABstar:9.2}, $\sqrt{S_{NN}}= 62.4$ GeV \cite{KABstar:62a}, $\sqrt{S_{NN}}= 130$ GeV \cite{KABstar:130a,KABstar:130b,KABstar:130c,KABstar:200a} and  200 GeV \cite{KABstar:200a,KABstar:200b,KABstar:200c}. 

To avoid possible biases we fit the particle ratios rather than the multiplicities. The best fit criterion is a minimality of $\chi^2 = \sum_i  \frac{(r^{theor}_i - r^{exp}_i)^2}{\sigma^2_i} $, where $r_i^{exp}$ is an experimental value of i-th particle ratio, $r_i^{theor}$ is our prediction and $\sigma_i$ is a total error of experimental value.

\begin{figure}[t]
\includegraphics[width=63mm]{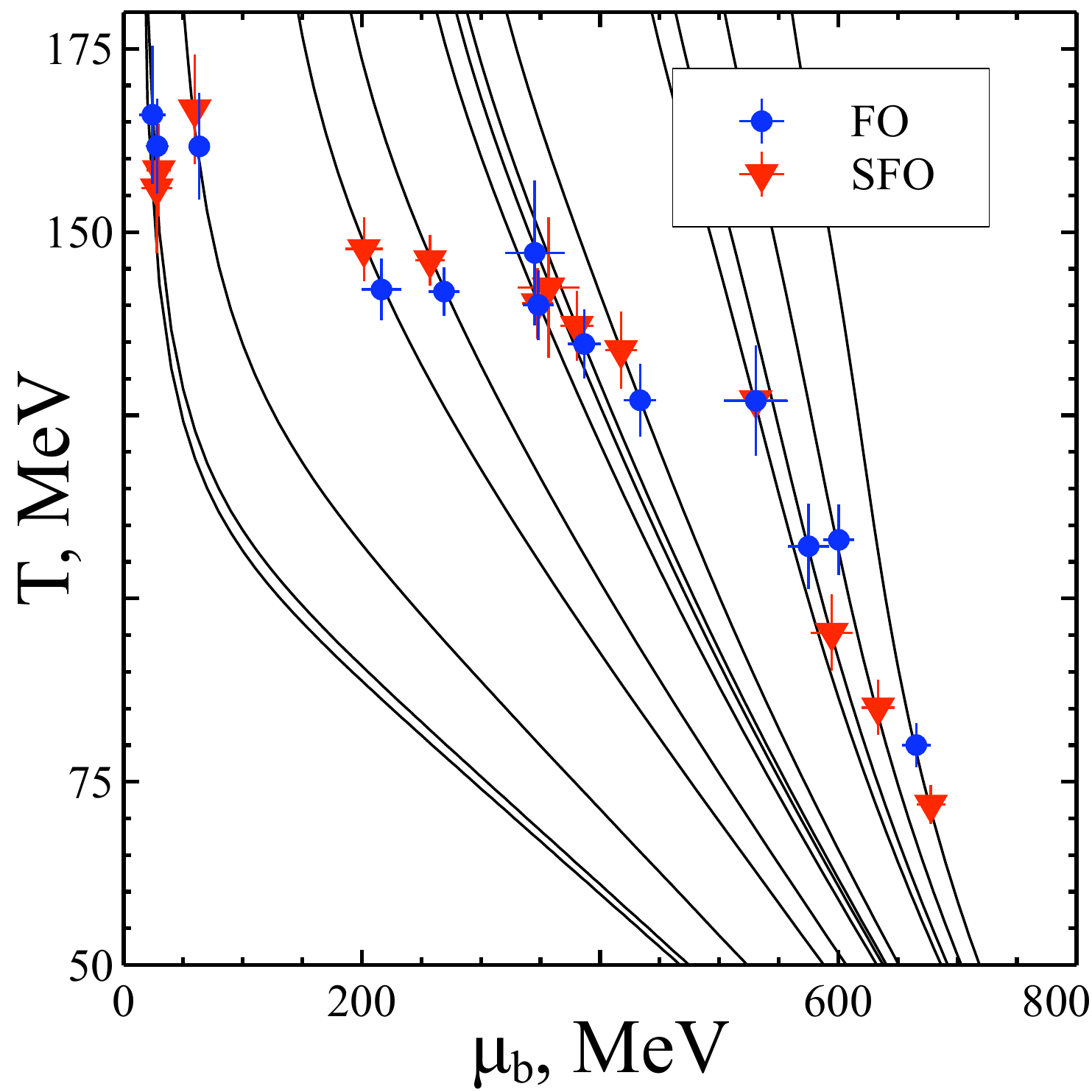}
\includegraphics[width=63mm]{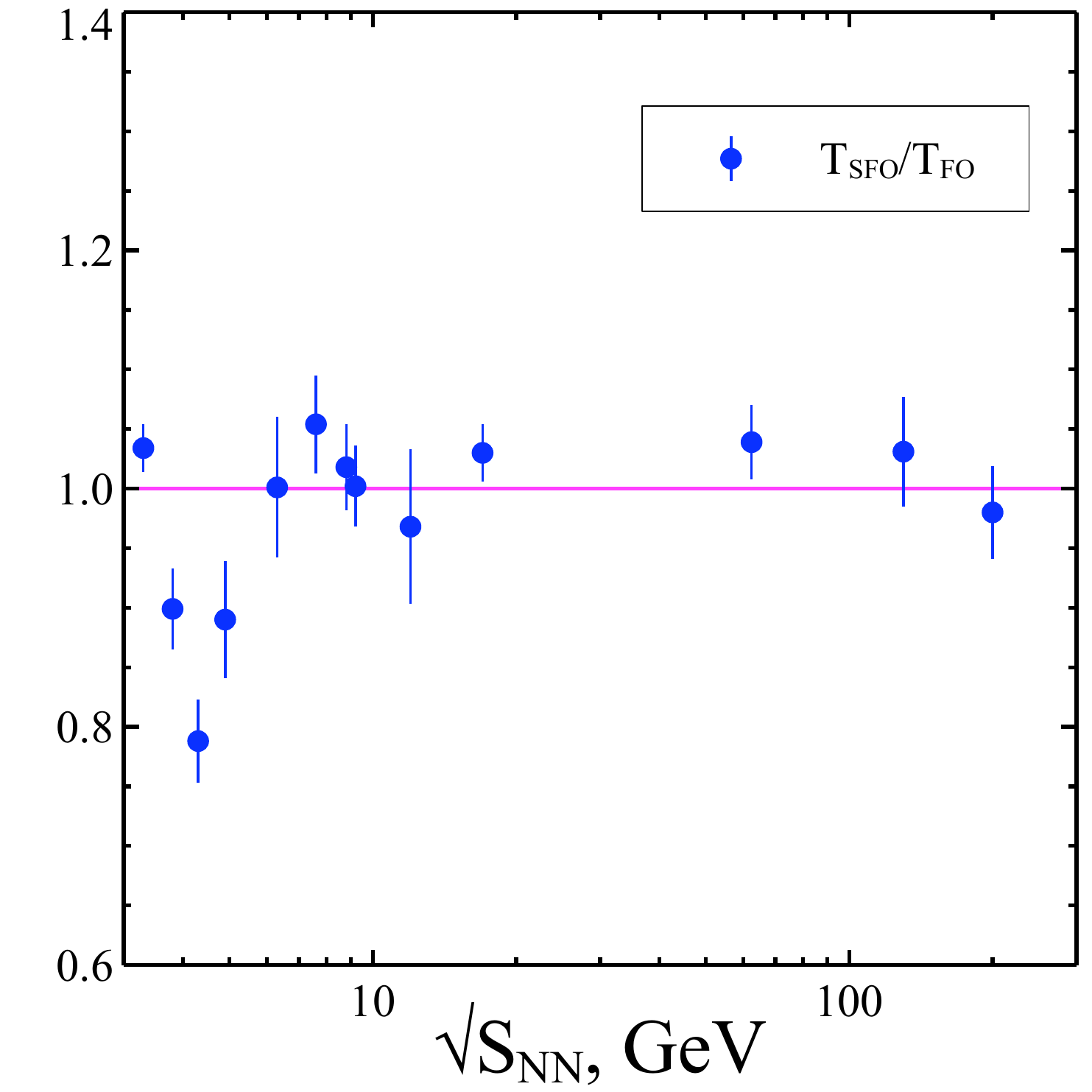}
\vspace*{-4.4mm}
\caption{Parameters of chemical freeze-outs in the model with two freeze-outs.   Upper panel: triangles correspond to the SFO, their coordinates are ($\mu_{B_{SFO}},\,T_{SFO}$), while circles  correspond to the FO and their coordinates are ($\mu_{B_{FO}},\,T_{FO}$). The curves  correspond to isentropic trajectories   ${s}/{\rho_B} = const$ connecting two freeze-outs.
Lower panel: $\sqrt{S_{NN}}$  dependence of the ratio of  the  SFO temperature to  the FO temperature.
}
\label{Fig:FO_SFO_param}
\end{figure}

\begin{figure}[t]
\includegraphics[width=63mm]{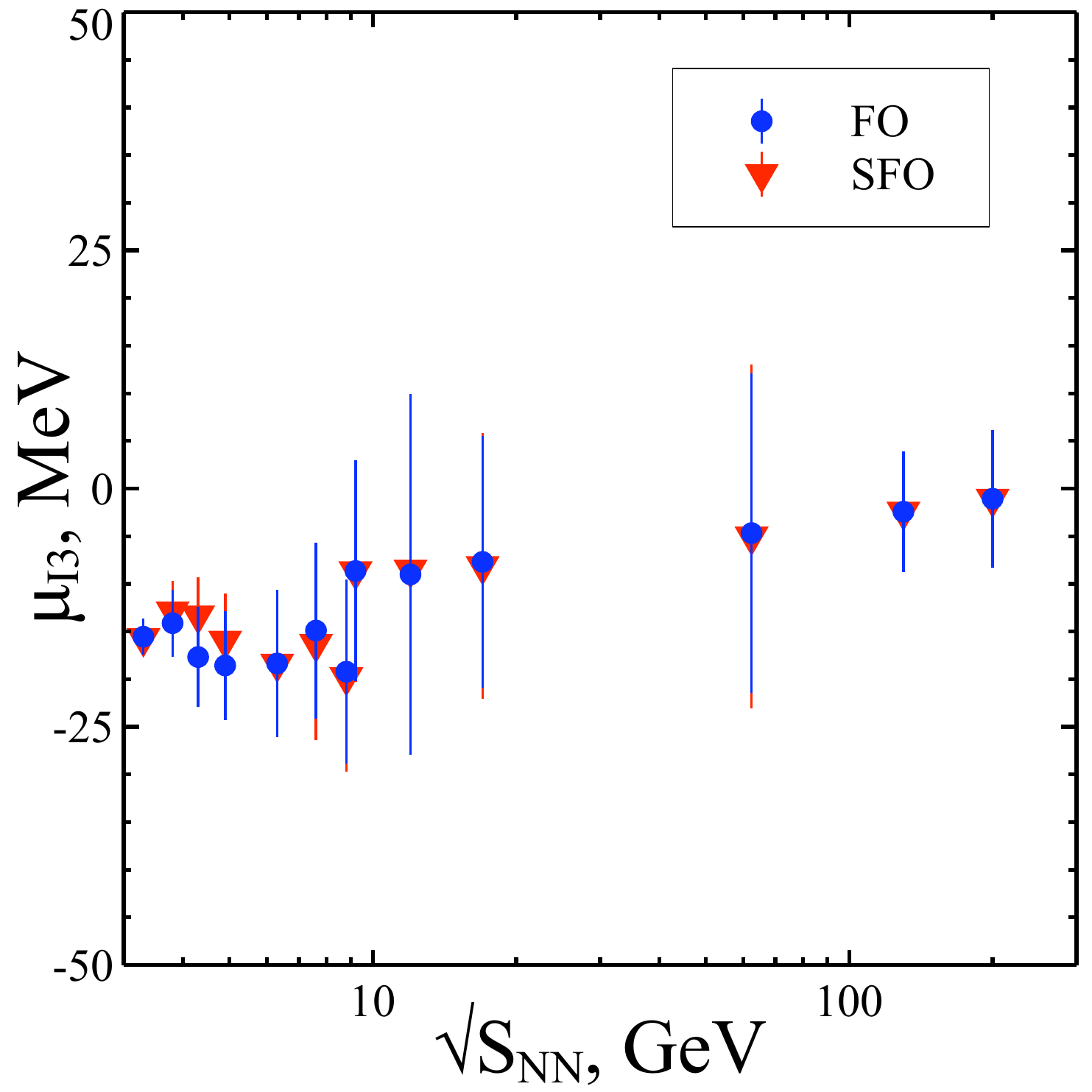}
\includegraphics[width=63mm]{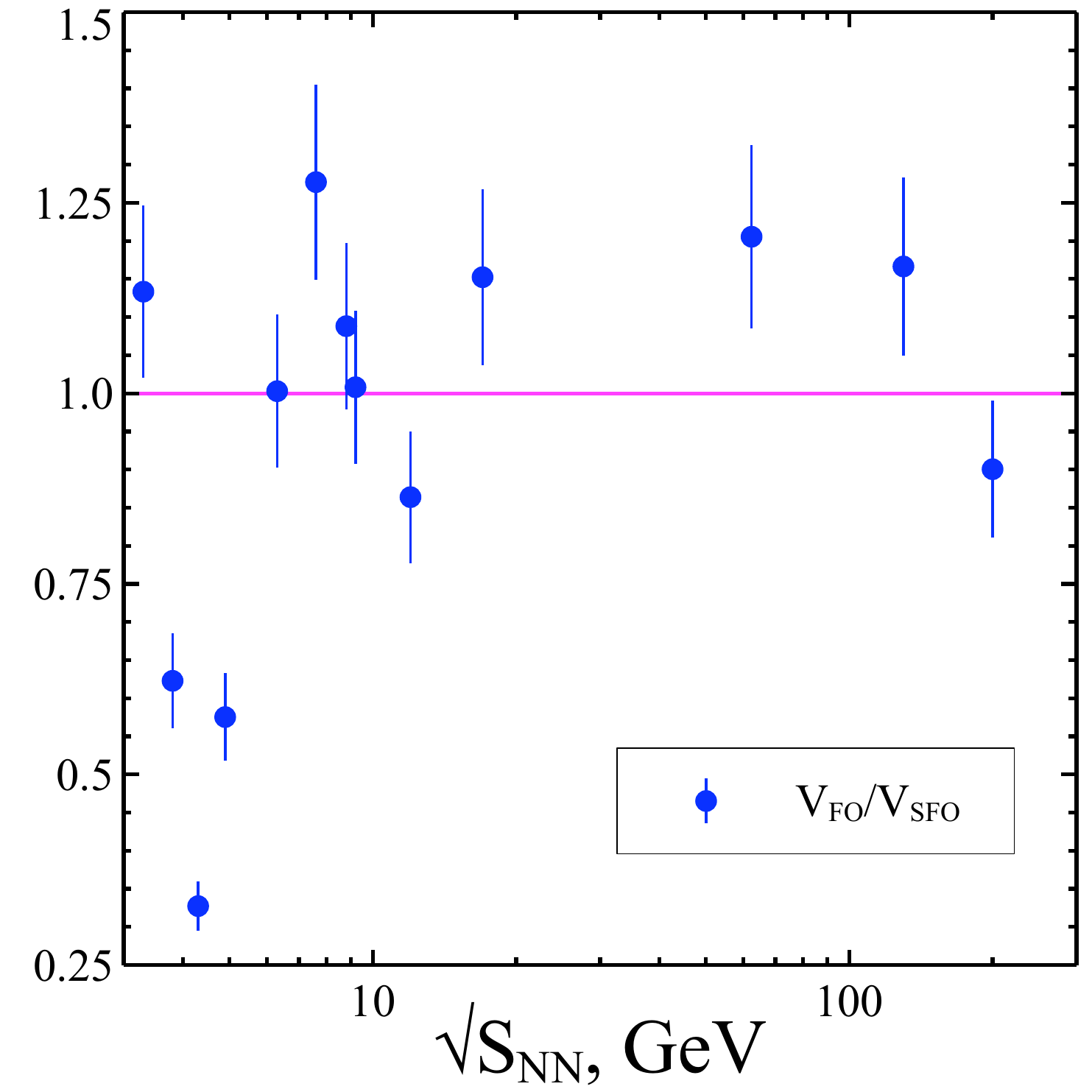}
\vspace*{-4.4mm}
\caption{Upper panel: $I_3$ chemical potential for  the FO (circles) and the SFO (triangles)
Lower panel: $\sqrt{S_{NN}}$  dependence of the ratio of the FO volume to the  SFO volume.
}
\label{Fig:muI3_Vrat}
\end{figure}

\begin{figure}[Htbp]
\vspace*{-4.0mm}
\includegraphics[height=63mm]{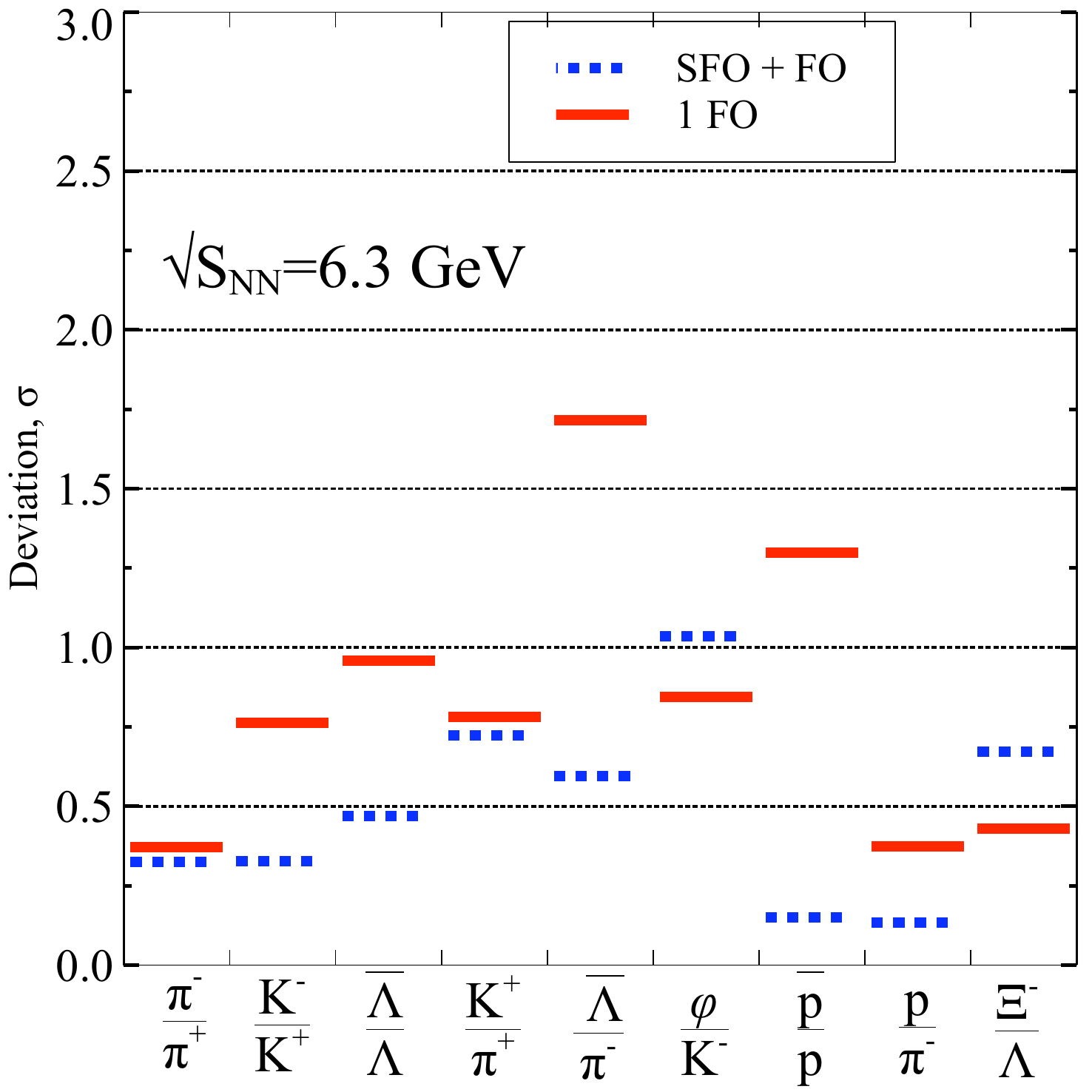}
\includegraphics[height=63mm]{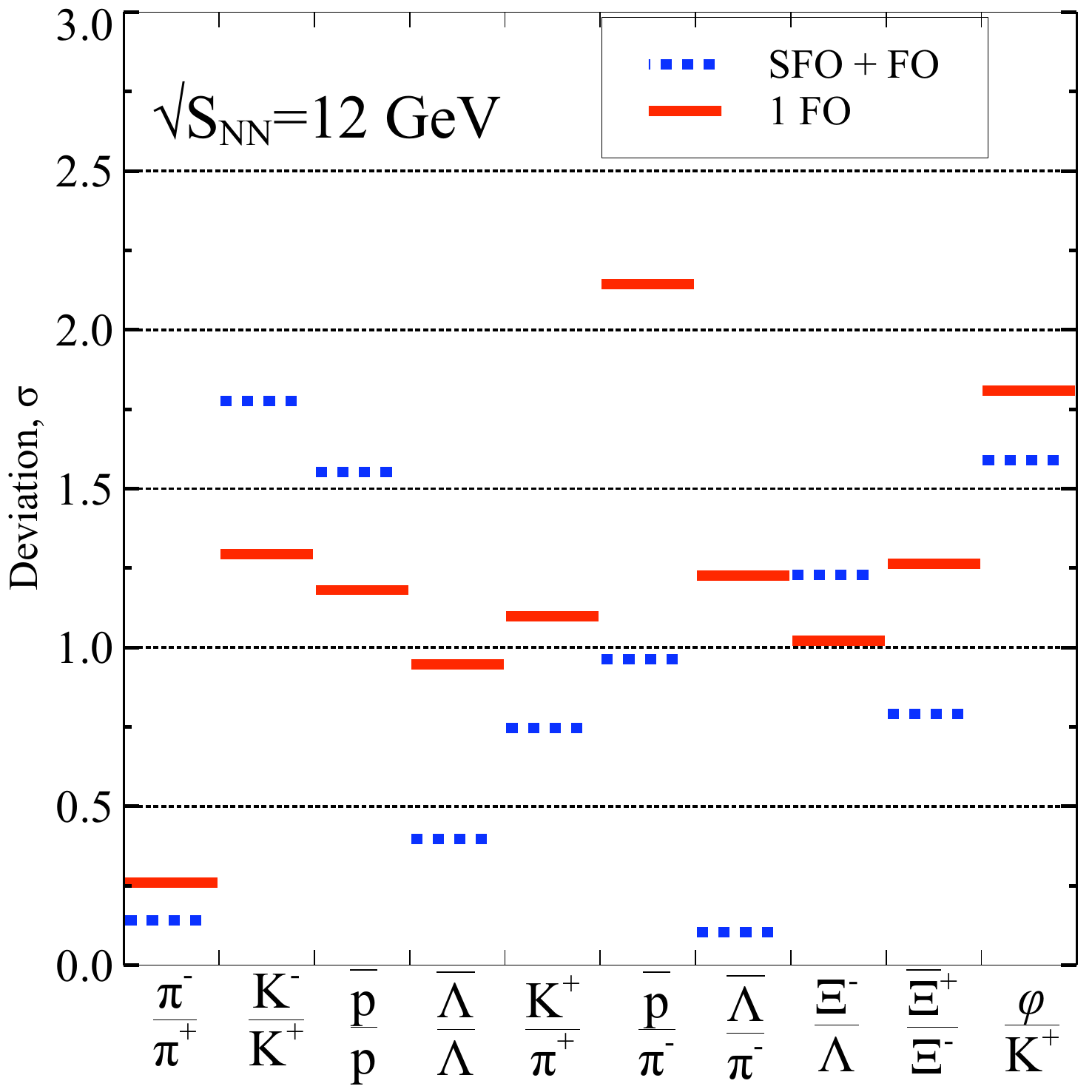}
\includegraphics[height=63mm]{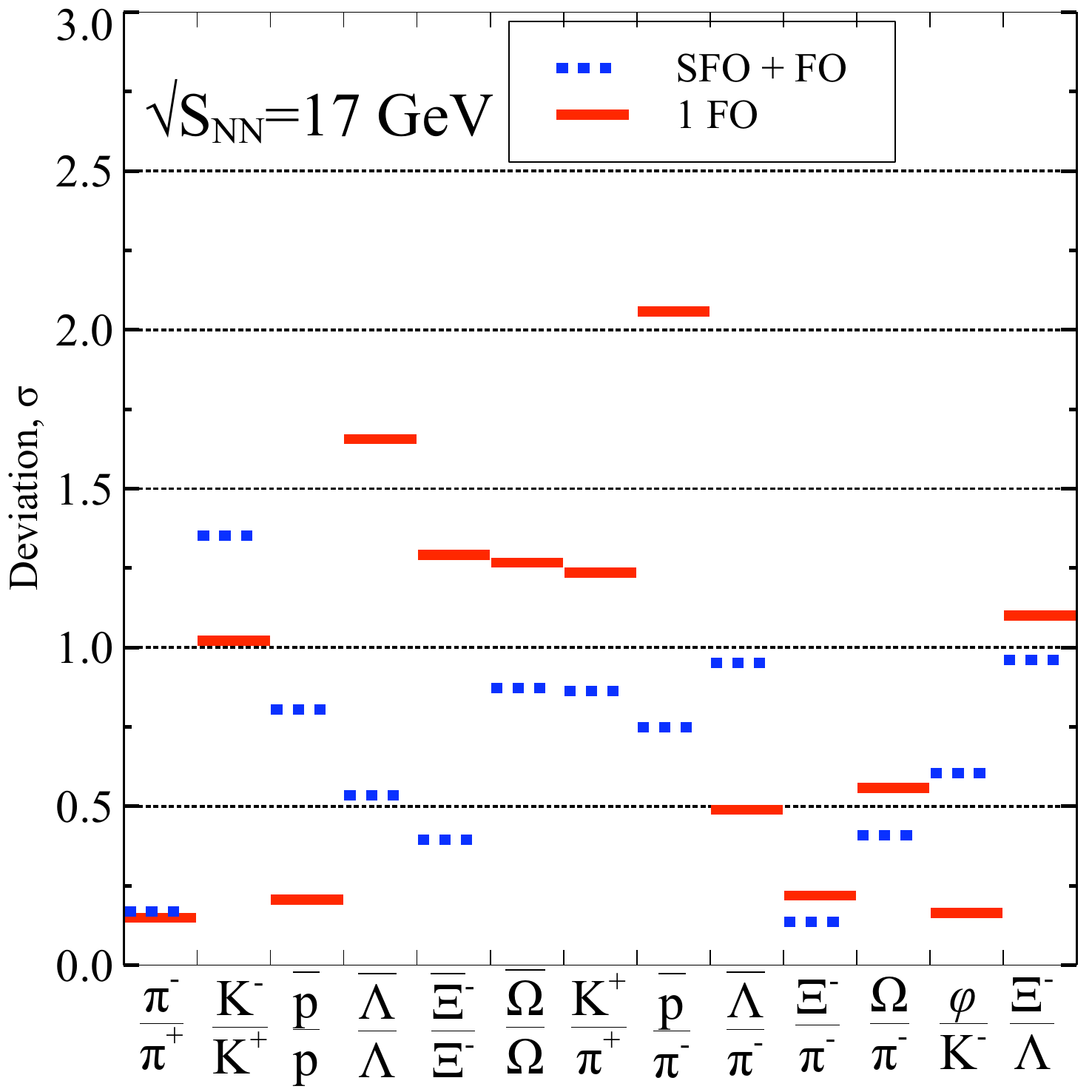}
\vspace*{-2.4mm}
 \caption{Relative deviation of theoretical description of ratios from experimental value in units of experimental error $\sigma$. The symbols on OX axis demonstrate the particle ratios. OY axis shows $\frac{|r^{theor} - r^{exp}|}{\sigma^{exp}}$, i.e. the modulus of  relative deviation for $\sqrt{S_{NN}}$ = 6.3, 12 and 17 GeV. The solid lines correspond to a model with one  chemical  freeze-out of all hadrons, while the dashed lines correspond to model with the  SFO.} 
  \label{Fig:dev_FO_SFO}
\end{figure}

{\bf Fit results.} The  FO and the SFO parameters are connected by  conservation laws (\ref{Eq:FO_SFO1}). Therefore, for the SFO  there is only one fitting parameter at each collision energy, namely $T_{SFO}$, while other parameters are found from the system  (\ref{Eq:FO_SFO1}). We study two things: behavior of parameters and what ratios are improved in the SFO approach compared to the case without SFO. First of all we found out that for SFO case $\chi^2/dof = 58.5/55 = 1.06$. At  $\sqrt{S_{NN}} = $ 2.7, 3.3, 3.8, 4.3 and 4.9 GeV the original  description obtained within the multicomponent model \cite{KABugaev:Horn2013} is very good and hence it has not improved significantly. Similar results are found at the highest RHIC energies $\sqrt{S_{NN}} >   62.4$ GeV. From Fig. \ref{Fig:FO_SFO_param} one can see that within these two energy domains the SFO temperatures demonstrate the largest deviations from the  FO temperature, although they do not exceed 20 \%. At intermediate  energies we see a systematic improvement of ratios description. Three plots corresponding to collision energies at which an improvement after SFO introduction is the most significant, $\sqrt{S_{NN}}$ = 6.3, 12 and 17 GeV,  are shown in Fig. \ref{Fig:dev_FO_SFO}. As one can see from Fig. \ref{Fig:dev_FO_SFO} for $\sqrt{S_{NN}}$ = 6.3, 12 and 17 GeV  the SFO approach improves description of all  ratios with more than one $\sigma$ deviation. For  $\sqrt{S_{NN}}$ = 6.3 GeV the SFO greatly   improves $\bar \Lambda/\pi^-$ and $\bar p/p$ ratios. For $\sqrt{S_{NN}}$ = 12 GeV  four ratios out of  eight   with more than one $\sigma$ deviation, namely  $K^+/\pi^+$, $\bar \Lambda/\Lambda$, $\bar \Lambda/\pi^-$ and  $\bar \Xi^+/\Xi^-$ are improved. The SFO approach allows us to significantly  improve the fit quality at $\sqrt{S_{NN}}$ = 17 GeV.  Fig. \ref{Fig:dev_FO_SFO} demonstrates that due to the SFO fit the six out of seven problematic ratios of the one freeze-out fit moved from the region of  deviation exceeding $\sigma$ to the region of deviations being smaller than $\sigma$. The most remarkable of them are $\bar p/\pi^-$, $\bar \Lambda/\Lambda$,  $\bar \Xi^-/\Xi^-$ and  $\bar \Omega/\Omega$. Thus, a separation of the FO and the SFO relaxes the strong connection  between the non-strange and strange baryons  and allows us not only to nicely describe the ratios of strange antibaryons to the same strange baryons, but also it allows us  for the first time to successfully reproduce the antiproton to pion ratio.

As we discussed above,  it is expected that  the SFO   occurs   earlier, when the system is smaller,  
and, hence,
 $V_{SFO} < V_{FO}$ or $\frac{V_{FO}}{V_{SFO}} > 1$. In the Fig. \ref{Fig:muI3_Vrat} one can see that this is, indeed, the case for most values  of  collision energy, but at low energies our expectation does not come true. One possible  formal   reason is the same as  for  an unexpected behavior of $\frac{T_{SFO}}{T_{FO}}$ (see Fig. \ref{Fig:FO_SFO_param}): at this energy range  the number of data points is just slightly larger than the  number of fitting  parameters and because of that  at low  energies of collisions the fit quality is very good without assumption of two freeze-outs.  There might be also a physical reason for such a behavior, namely at  low collision energies 
 the freeze-out occurs at large baryonic densities which may essentially affect the in medium cross-sections of  the reactions with strangeness exchange due to additional attraction and, therefore, such reactions do not freeze-out  earlier than other reactions. 

Finally, we would like to suggest a generalization of  the double freeze-out HRGM that will be able to ultimately improve the  description of  multiplicities. The first step is to identify the hadronic reactions that freeze-out noticeably earlier or later than  most of the  others. This should be done separately for each collision energy, since for different energies the reaction  cross-sections, the particle concentrations and the fireball expansion rate are different. Such reactions may be identified using the  system   (\ref{eq:equil})  or by running the transport model code  and counting for the reaction rates versus time. If such reactions exist, then  their  separate freeze-out  should  be considered. It is clear that the conservation laws between the freeze-outs may be different depending on what reactions are switched off. For instance, if all reactions with the $\Omega$ hyperon are frozen, then the  conservation law of the number of $\Omega$ hyperons  should  be introduced. Probably, the charmed particles are good candidates for 
the separate freeze-out.

\section{Conclusions}

Here we  thoroughly  discussed 
an  assumption that in heavy ion collisions the strangeness exchange reactions  may freeze-out earlier.
Using such an assumption  we constructed a modification of the HGRM with two freeze-outs, connected with the conservation laws. One freeze-out  corresponds to all strange particles and another freeze-out is for all non-strange ones. The conservation laws allow us for each collision energy  to get  just one additional fitting parameter 
compared to the HRGM with a simultaneous  chemical freeze-out of all hadrons.
 We have shown that such a model describes 111 independent  hadron ratios measured at $\sqrt{S_{NN}}$ = 2.7 - 200 GeV even better than the most elaborate  version of the HRGM with a single freeze-out   ($\chi^2/dof $ = 1.06 for the model with two freeze-outs  versus 1.16 for one freeze-out).

We suggest to go even further: for each collision energy to separately identify the processes which  freeze-out at considerably different time than all the other and  to construct a corresponding HRGM  with two freeze-outs. Identification of  such reactions can be done using the transport models.

{\bf Acknowledgments.}
The authors are thankful to to P. Huovinen for fruitful discussions.
D.R.O. acknowledges funding of a Helmholtz Young Investigator Group VH-NG-822
from the Helmholtz Association and GSI, and   thanks HGS-HIRe for a support.
A.I.I.  and  K.A.B.  acknowledge  a  support 
of  the Fundamental Research State Fund of Ukraine, Project No F58/04.
Also K.A.B.   acknowledges  a partial support provided by the Helmholtz 
International Center for FAIR within the framework of the LOEWE 
program launched by the State of Hesse.

\end{document}